\documentclass{iau}
\usepackage{graphicx}

\newcommand{\Mbh}{M_{\bullet}}

\newcommand{\Mo}{M_{\odot}}

\mathchardef\mhyphen="2D

\newcommand{\yr}{{\,\rm yr}}

\newcommand{\Myr}{{\,\rm Myr}}

\newcommand{\pc}{\,\mathrm{pc}}

\newcommand{\kpc}{\,\mathrm{kpc}}

\title[Young Stars at the Galactic Center]
{On the Origin of Young Stars  \\ at the Galactic Center}

\author[Madigan et al.]{Ann-Marie Madigan$^{1,2,3}$, Oliver Pfuhl$^{4}$, Yuri Levin$^{3,5}$, Stefan Gillessen$^{4}$, Reinhard Genzel$^{4,6}$ and Hagai B. Perets$^{7}$}

\affiliation{$^1$Astronomy Department and Theoretical Astrophysics Center, University of California, Berkeley, CA 94720, USA \\$^2$Einstein Postdoctoral Fellow; email: {\tt ann-marie@astro.berkeley.edu} \\$^3$Leiden Observatory, Leiden University, P.O. Box 9513, NL-2300 RA Leiden, The Netherlands \\$^4$Max-Planck Institut f\"ur Extraterrestrische Physik, 85748 Garching, Germany \\$^5$School of Physics, Monash University, Clayton, Victoria 3800, Australia \\$^6$Department of Physics, University of California, Berkeley, CA 94720, USA \\$^7$Physics Department, Technion - Israel Institute of Technology, Haifa, Israel 32000}

\pubyear{2013}
\volume{303} 

\jname{ The Galactic Center: \\Feeding and Feedback in a Normal Galactic Nucleus}
\editors{}
\begin{document}

\bibliographystyle{apj} 

\makeatletter

\let\jnl@style=\rm
\def\ref@jnl#1{{\jnl@style#1}}

\def\aj{\ref@jnl{AJ}}                   
\def\actaa{\ref@jnl{Acta Astron.}}      
\def\araa{\ref@jnl{ARA\&A}}             
\def\apj{\ref@jnl{ApJ}}                 
\def\apjl{\ref@jnl{ApJ}}                
\def\apjs{\ref@jnl{ApJS}}               
\def\ao{\ref@jnl{Appl.~Opt.}}           
\def\apss{\ref@jnl{Ap\&SS}}             
\def\aap{\ref@jnl{A\&A}}                
\def\aapr{\ref@jnl{A\&A~Rev.}}          
\def\aaps{\ref@jnl{A\&AS}}              
\def\azh{\ref@jnl{AZh}}                 
\def\baas{\ref@jnl{BAAS}}               
\def\bac{\ref@jnl{Bull. astr. Inst. Czechosl.}}
\def\caa{\ref@jnl{Chinese Astron. Astrophys.}}
\def\cjaa{\ref@jnl{Chinese J. Astron. Astrophys.}}
\def\icarus{\ref@jnl{Icarus}}           
\def\jcap{\ref@jnl{J. Cosmology Astropart. Phys.}}
\def\jrasc{\ref@jnl{JRASC}}             
\def\memras{\ref@jnl{MmRAS}}            
\def\mnras{\ref@jnl{MNRAS}}             
\def\na{\ref@jnl{New A}}                
\def\nar{\ref@jnl{New A Rev.}}          
\def\pra{\ref@jnl{Phys.~Rev.~A}}        
\def\prb{\ref@jnl{Phys.~Rev.~B}}        
\def\prc{\ref@jnl{Phys.~Rev.~C}}        
\def\prd{\ref@jnl{Phys.~Rev.~D}}        
\def\pre{\ref@jnl{Phys.~Rev.~E}}        
\def\prl{\ref@jnl{Phys.~Rev.~Lett.}}    
\def\pasa{\ref@jnl{PASA}}               
\def\pasp{\ref@jnl{PASP}}               
\def\pasj{\ref@jnl{PASJ}}               
\def\rmxaa{\ref@jnl{Rev. Mexicana Astron. Astrofis.}}%
\def\qjras{\ref@jnl{QJRAS}}             
\def\skytel{\ref@jnl{S\&T}}             
\def\solphys{\ref@jnl{Sol.~Phys.}}      
\def\sovast{\ref@jnl{Soviet~Ast.}}      
\def\ssr{\ref@jnl{Space~Sci.~Rev.}}     
\def\zap{\ref@jnl{ZAp}}                 
\def\nat{\ref@jnl{Nature}}              
\def\iaucirc{\ref@jnl{IAU~Circ.}}       
\def\aplett{\ref@jnl{Astrophys.~Lett.}} 
\def\apspr{\ref@jnl{Astrophys.~Space~Phys.~Res.}}
\def\bain{\ref@jnl{Bull.~Astron.~Inst.~Netherlands}} 
\def\fcp{\ref@jnl{Fund.~Cosmic~Phys.}}  
\def\gca{\ref@jnl{Geochim.~Cosmochim.~Acta}}   
\def\grl{\ref@jnl{Geophys.~Res.~Lett.}} 
\def\jcp{\ref@jnl{J.~Chem.~Phys.}}      
\def\jgr{\ref@jnl{J.~Geophys.~Res.}}    
\def\jqsrt{\ref@jnl{J.~Quant.~Spec.~Radiat.~Transf.}}
\def\memsai{\ref@jnl{Mem.~Soc.~Astron.~Italiana}}
\def\nphysa{\ref@jnl{Nucl.~Phys.~A}}   
\def\physrep{\ref@jnl{Phys.~Rep.}}   
\def\physscr{\ref@jnl{Phys.~Scr}}   
\def\planss{\ref@jnl{Planet.~Space~Sci.}}   
\def\procspie{\ref@jnl{Proc.~SPIE}}   

\let\astap=\aap
\let\apjlett=\apjl
\let\apjsupp=\apjs
\let\applopt=\ao
\makeatother

\maketitle

\begin{abstract}

The center of our galaxy is home to a massive black hole, SgrA*, and a nuclear star cluster containing stellar populations of various ages. While the late type stars may be too old to have retained memory of their initial orbital configuration, and hence formation mechanism, the kinematics of the early type stars should reflect their original distribution. 
In this contribution we present a new statistic which uses directly-observable kinematical stellar data to infer orbital parameters for stellar populations, and is capable of distinguishing between different origin scenarios. We use it on a population of B-stars in the Galactic center that extends out to large radii ($\sim0.5 \pc$) from the massive black hole. We find that the high $K$-magnitude population ($ \lesssim 15 \Mo$) form an eccentric distribution, suggestive of a Hills binary-disruption origin. 

\end{abstract}

\firstsection 
\section{Where do the young stars come from?}

The nuclear star cluster in the central few parsecs of our Galaxy contains a rich concentration of young massive stars (see review by Jessica Lu in this volume). This is surprising as stars must overcome a hostile tidal field to form close to a massive black hole. In this paper we focus on their origin. There appear to be at least two dynamically distinct populations of young stars: 1) a clockwise rotating disk with very massive stars (including the most massive OB and Wolf-Rayet types) which is thought to have formed about $\sim 4-6 \Myr$ ago from a fragmenting gaseous disk, 2) a cluster of less massive B-stars which appear more isotropically distributed and do not respect the disk inner edge. The B-stars with projected radii of less than one arcsecond ($\sim 0.04 \pc$ at $8.3 \kpc$) are commonly referred to as the `S-stars'. Their ages are not well constrained; they may well be co-eval with the $4-6 \Myr$ population but could also be much older, up to a few $100 \Myr$. It is likely that they originated via the tidal disruption (due to the massive black hole) of stellar binaries, a mechanism which deposits captured stars on highly-eccentric orbits whilst flinging their companions out at high velocity (\cite[Hills 1988]{Hil88}). The question we're interested in is whether the B-stars outside the central arcsecond are part of the episode of star formation which produced the clockwise disk structure, a remnant of a former nuclear disk formation event, or simply a continuation of the S-stars, each one a captured component of a tidally disrupted binary.

The key to distinguishing between formation scenarios is stellar orbital eccentricity. Stellar orbital eccentricities should be low for a disk formation origin, particularly if the gas disk from which the stars formed had time to circularize before fragmentation, and very high for the binary disruption mechanism ($e \gtrsim 0.97$). As dynamical relaxation times increase with radius from the massive black hole, we expect B-stars at large semi major axes to retain their initial orbital eccentricities over timescales longer than their main sequence lifetimes (\cite[Perets \& Gualandris 2010]{Per10}). Therefore, by measuring the eccentricities of the B-stars at large radii, we should know which origin scenario is correct. Stars with large semi major axes have large orbital periods ($P (a \sim 0.1 - 0.5 \pc) \sim 10^{3-4} \yr$) however, and it is therefore difficult to determine their dynamical accelerations from astrometric data, from which orbital parameters are derived. To circumvent this, we have developed a statistical method using sky positions and proper motions to infer the orbital eccentricities, not of individual stars, but of a population of stars as a whole. It is particularly sensitive to high eccentricity populations (hence called the $h$-statistic), suitable for the binary disruption model. For each star we calculate 
\begin{equation}
h = \frac{x v_y - y v_x }{\sqrt{G \Mbh p}},
\end{equation}
where $v_x,v_y$ are its right ascension and declination velocities at projected radius $p = \sqrt{x^2 + y^2}$ on the sky, and $\Mbh$ is the mass of the massive black hole. $h$ is $\sim\! 1, \sim\! 0, \sim\! -1$ depending on whether the star's orbit projected on the sky is mainly clockwise tangential, radial, or counter-clockwise tangential. Radial orbits are confined to low $|h|$-values. If a population of stars is isotropically distributed, the presence of many low $|h|$-values directly implies a population with high orbital eccentricities. 

We simulate B-stars in two formation scenarios, one based on the binary disruption model in which the B-stars have high-eccentricity orbits, and another based on a disk formation model in which B-stars have lower eccentricities. We evolve their orbits for $6 \Myr$ using an $N$-body integrator (described in \cite[Madigan et al. 2011]{Mad11b}) in a Galactic center potential which includes a massive black hole, nuclear star cluster and massive clockwise disk. We compare the results using the new $h$-statistic with VLT adaptive optics observations in different $K$-band ($2.17\, \rm \mu m$)  magnitude ranges. This short contribution reports the main results. A more detailed analysis is presented in \cite[Madigan et al. 2014]{Mad14}. 

\section{Comparison between simulations and data}

\begin{figure}[b]
\begin{center}
 \includegraphics[angle = -90,width=3.4in]{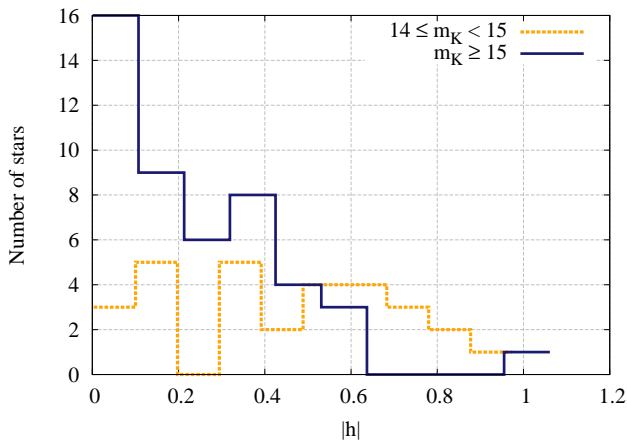} 
 \caption{Histogram of $|h|$-values for stars with $K$-magnitude $14 \leq m_K < 15$ and $m_K \geq 15$. Data are published in \cite[Madigan et al. 2014]{Mad14}.}
   \label{fig1}
\end{center}
\end{figure}

\begin{figure}[b]
\begin{center}
 \includegraphics[angle = -90,width=3.4in]{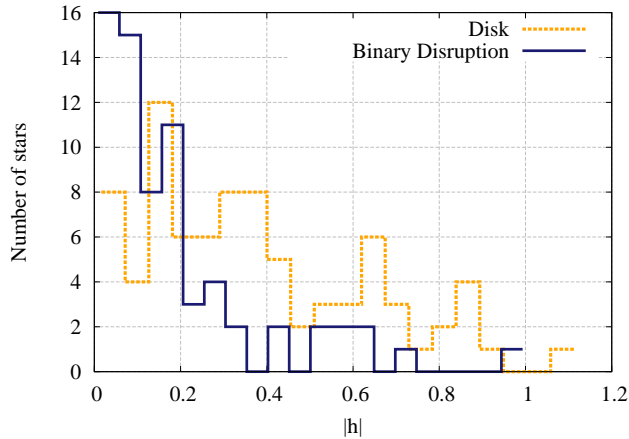} 
 \caption{Histogram of $|h|$-values of stars in simulations, corrected for completeness.}
    \label{fig2}
\end{center}
\end{figure}

In Fig.\,\ref{fig1} we show histograms of $|h|$-values of stars in different $K$-magnitude ranges. Mapping from $K$-band magnitude to stellar mass is dependent on models of stellar evolution and atmospheres, but in general lower magnitude stars convert to higher masses and vice versa. For comparison we plot histograms of $|h|$-values of simulated B-stars in a binary disruption origin and a disk formation origin, corrected for observational completeness, after $t = 6 \Myr$ in Fig.\,\ref{fig2}. We find evidence for a population of B-stars, $m_K \geq 15$, on low angular momentum orbits, i.e., with high orbital eccentricity, within the central half parsec of the Galactic center. These stars may be the captured components of tidally disrupted stellar binaries originating outside the central few parsecs of the Galactic center (\cite[Perets et al., 2007]{Per07}). Lower magnitude B-stars, $14 \leq m_K < 15$, match better to a lower-eccentricity disk formation origin. The short main sequence lifetimes of these high-mass stars suggest that they are from the same star formation episode as the massive, young clockwise rotating disk.

\section{Implications}

If a significant number of B-stars at the radii of the young clockwise disk originated, as our results suggest, via the tidal disruption of binaries there are several consequences. 

\begin{enumerate}

\item In determining the initial mass function for the young clockwise disk, a $K$-band luminosity function is typically constructed from stars born in that epoch of star formation. A luminosity-mass relation from stellar evolution and atmosphere models can then be used to convert from observed magnitudes to initial stellar masses. Current estimates find a top-heavy initial mass function for the disk (\cite[Bartko et al. 2010, Do et al. 2013, Lu et al. 2013] {Bar10, Do13, Lu13}). However, if many B-stars originate from the binary disruption mechanism, they should not be included in the $K$-band luminosity function. The initial mass function of the young clockwise disk in this case must be even more top-heavy than previously reported. 

\item 
The binary disruption mechanism deposits stars on low angular momentum (near-radial) orbits centered on the massive black hole. This provides us with a natural experiment with which to probe the underlying stellar mass distribution, the majority of which is too under-luminous to be observable. The rate at which the near-radial orbits relax or diffuse in angular momentum depends on the stellar mass distribution as a function of radius from the massive black hole. The $h$-values of the B-stars as a function of their radii can put constraints on the underlying mass density profile of the stellar potential. The data are tentatively suggestive of a cusp rather than a core in the Galactic center; a more detailed analysis and a larger data sample are necessary to confirm this (\cite[see Madigan et al. 2014]{Mad14}).

\end{enumerate}

\acknowledgments{A.-M. M is supported by NASA Einstein Fellowship grant PF2-130095.}

\end{document}